**Observation of three-dimensional behavior in surface states of bismuth nanowires and the evidence for bulk Bi charge fractionalization.**

T.E. Huber,[1] A. Nikolaeva,[2] L. Konopko,[2] and M.J. Graf [3]

[1] Howard University, Washington, DC 20059-0001

[2] Academy of Sciences, Chisinau, Moldova and International Laboratory of High Magnetic Fields and Low Temperatures, Wroclaw, Poland.

[3] Department of Physics, Boston College, Chestnut Hill MA 02467

**ABSTRACT**

Whereas bulk bismuth supports very-high mobility, light, Dirac electrons and holes in its interior, its boundaries support a layer of heavy electrons in surface states formed by spin orbit interaction in the presence of the surface electric field. Small diameter $d$ trigonal Bi nanowires (30 nm < $d$ < 200 nm) were studied via magnetotransport at low temperatures and for fields up to 14 T in order to investigate the role of surfaces in electronic transport. A two-dimensional behavior was expected for surface charges; however we found instead a three-dimensional behavior, with a rich spectrum of Landau levels in a nearly spherical Fermi surface. This is associated with the long penetration length of surface states of trigonal wires. The prospect of the participation of surface transport and surface-induced relaxation of bulk carriers in the electronic properties of macroscopic samples is evaluated. We show that recent observations of magnetoquantum peaks in the Nernst thermopower coefficient, attributed to two-dimensional electron gas charge fractionalization, can be more plausibly interpreted in terms of these surface states.

Semimetallic bismuth (Bi) is one of the most studied conductors because of its extreme electronic properties such as very high electron mobility and Hall coefficients. Bi has a Fermi surface (FS) consisting of small hole and electron pockets at T- and L-points, respectively,[Fig.1(a)] and therefore the Fermi wavelength $\lambda_F$ is very long (about 50 nm). The magnetic energy, that is inversely proportional to $\lambda_F$, is sufficiently large in comparison with the Fermi energies to depopulate Bi of holes in magnetic field (*H*) as small as 9 T. The high-*H* state has been studied experimentally[1] and the results were well described by the model of holes and Dirac electrons with strong spin-orbit interaction presented in Ref. 2. Therefore, the interpretation by Behnia, Balicas and Kopelevich[3] of their measurements of Nernst thermopower and Hall effect of bulk Bi in terms of charge fractionalization, a collective quantum state, was unexpected because it would involve new quasiparticles. Subsequently, Li *et al* [4] presented a study of the magnetic properties that was largely confirmatory of the results by Behnia *et al*. Clearly, the observation in bulk Bi [5] of two-dimensional (2D) quantum phenomena [6], as observed in 2D electron gases [7] and has been discussed theoretically for three-dimensional (3D) gases[8] and for the Bi boundary [9] is very exciting. But Refs. 3 and 4 do not address the main issues raised by their interpretation, that is, what is the nature of the quasiparticles and the origin and dimensionality of the phase underlying the effect.

Because electrons and holes in bulk Bi exhibit very long relaxation times, the electronic mean free path in bulk Bi is several millimeters long,[10] significantly longer than the size of the samples in the experiments. Therefore it can be presumed that the boundary plays an important role in the energy relaxation of the charge carriers. Since the thermoelectric potentials, as well as the electric potentials associated with wrapping (surface) currents, are determined by relaxation processes,[11] surface relaxation effects will play a key role in low temperature experiments involving Bi, and in particular in the experiments in Refs. 3 and 4. Direct observations via angle resolved photoemission spectroscopy (ARPES) show that semi-infinite Bi surfaces support a large density of surface states that are populated by electrons of density $\Sigma \sim 2-5 \times 10^{12}$ cm$^{-2}$, effective mass $m_\Sigma \sim 0.2$ and exhibit an anisotropic penetration length.[12] By applying ARPES to ultrathin Bi films, Hirahara *et al*,[13] provided evidence of the bulk-like states and boundary states in Bi and demonstrated that they are hybridized. Hybridization could be a primary mechanism for energy relaxation of bulk carriers. However ARPES measurements in Ref. 13 did not measure transport properties or the effect of magnetic fields, and therefore the connection between relaxation processes and the results of Refs. 3 and 4 is not clear. Here we report magnetotransport measurements in high mobility 30-, 50- and 200-nm diameter Bi nanowire arrays, for which we observe a spectrum of resonances of bulk-like quasiparticles and surface carriers as a function of magnetic field up to 16 T. Our results address directly the matter of the participation of the boundary in the transport properties of solid samples. Preliminary studies have been presented.[14-16] Because of their special periodicity in 1/$H$, we argue that we observe resonances that are caused by Landau states (LS) in a spherical Fermi surface formed in the body of the nanowires but probably in close proximity to the boundary, from the surface states in the nanowire boundary. The surface states of Bi surfaces perpendicular to the trigonal direction have long penetration length as shown by ARPES. We contend that the nanowires still exhibit the same physics that semi-infinite surfaces do because our observed spectrum of resonances (the location of the peaks in 1/$H$) of the boundary's LS and the resonances observed by Behnia are remarkably similar. This then calls into question the interpretation of results in terms of fractionalization presented in Ref. 3.

Depending upon the diameter *d*, nanowires may present extreme behaviors because of quantum confinement effects [17] and the large surface-to-volume ratio. While the LS spectrum of electrons and holes in the large diameter nanowires, apart small quantum confinement shifts, is analogous to that for bulk Bi, and the electrons and holes are the majority carriers, the very small diameter nanowires ($d < \lambda_F$) realize the quantum limit without magnetic fields. For these nanowires, the quantum confinement energy $E_c = \hbar^2 \pi^2 / 2 m^* m_e d^2$, where $m_e$ is the electron mass and $m^*$ is the effective mass in units of $m_e$, exceeds the electron-hole overlap energy $E_0$. Therefore *n* and *p*, electron and hole densities are decreased critically below the bulk equilibrium values ($n_0 = p_0 = 3 \times 10^{17}$ cm$^3$), resulting in a semimetal(SM)-to-semiconductor (SC) transition (SMSC).[18] The SMSC critical diameter for holes and electrons for our nanowires ($m^* = 0.065$) is around 50 nm. Accordingly, it is expected that 30-nm wires are SC, 200-nm wires are SM, and 50-nm wires are at the SMSC transition. Given the reduction in the bulk electron *n* and hole *p* densities relative to the surface carrier concentration $\Sigma$, one expects the surface carriers to become a clear majority in nanowires with diameters below 100 nm at low temperatures; for example, the ratio of surface carriers per unit length to bulk electrons or holes per unit length is 15 for 50-nm wires. This ratio is even larger for nanowires on the SC side of the SMSC transition.

Only high mobility samples exhibit the sharp resonances that we identify as LS phenomena. Oxidation, impurity diffusion from electrodes and solder, and the low-melting point and sensitivity to mechanical strains, make it difficult to fabricate high mobility low-dimensional Bi samples suitable for electronic transport measurements. Here, we employ high mobility intrinsic nanowire samples, namely arrays of trigonal nanowires (the orientation is shown in Fig. 1(a)) of diameters ranging between 30 and 200 nm prepared by high pressure infiltration of pure Bi (99.999%) into nanochannel alumina templates.[15] Our two-probe electrical contact technique consists of attaching Cu wire electrodes to both sides of the Bi nanowire composite by using silver epoxy contacts. We will show below that these samples support LS. The LS levels inverse-field oscillatory period is $\Delta(1/H)=e/(cA_e)$, where $A_e$ is any extremal cross-sectional area of the Fermi surface perpendicular to the magnetic field.[19] As the magnetic field is increased, the energy of the Landau states increases, and when the energy of a particular level equals the Fermi energy, this level becomes depopulated. The relaxation time for electron scattering is temporarily increased at this field value, giving rise to a dip in the magnetoresistance. For an ellipsoidal FS, the carrier density is $p = (1/3\pi)A_\perp A_{//}^{1/2}$ where // and $\perp$ refer to the directions along the major and minor axis. The effective mass ($dA/dH$) and relaxation time can be extracted from the oscillatory magnetoresistance (MR), or Shubnikov-de Haas (SdH) effect. Zero-field resistance and MR measurements were made in two separate laboratories: (i) Boston College in a cryostat operating over the range 1.8 K < $T$ < 300 K and $H$ < 9 T and (ii) International Laboratory of High Magnetic Fields and Low Temperatures (Wroclaw, Poland) in a cryostat operating between 1.8 K < $T$ < 300 K and $H$< 14 T, and which allowed for *in-situ* sample rotation around two axes. This makes possible the measurement of the longitudinal MR (LMR), transverse MR (TMR), and intermediate orientations $\theta$ of the nanowires with respect to the magnetic field; the mesh was limited to $15^0$ due to magnet time limitations. In order to successfully interpret LS data for small-diameter nanowires through the SdH method one needs to consider the electronic transport regime (ballistic versus diffusive) and Chambers's effect.[20] The Chambers magnetic field, $H_c$, for which the Landau level orbit diameter equals the wire diameter, is given by $H_c = 2\hbar c k_F / ed$, where $k_F$ is the actual Fermi wave vector, $\hbar$ is Planck's constant, $c$ the speed of light, and $e$ the electronic charge. For high purity samples it is expected that LMR < TMR when $H > H_c$ because at high magnetic fields, the carriers avoid collision with the walls when the magnetic field is oriented along the wire direction. Our samples indeed satisfy this condition.

Figure 2 shows $d(MR)/dH$ versus $1/H$ for the nanowires in our study at low temperatures for several illustrative values of $\theta$. In the 200-nm case, the Dirac electron and hole LS extrema that we observe can be indexed in terms of quantum number and spin, as shown by Smith *et al*.[21] According to this correspondence the strong LMR feature at 0.125 $T^{-1}$, at the end of our observation range (9 T) is the h1⁻, the singlet at the hole quantum limit (there are no hole LS eigenstates at higher magnetic fields). The peak indexed h2⁻ is the 2⁻,0⁺ hole doublet. It is observed at 0.255 $T^{-1}$ in 200-nm Bi nanowires, at 0.24 $T^{-1}$ in bulk Bi with MR, and it is also reported in [3] The location of the peaks in the macroscopic bulk can be calculated from the Fermi energy of holes, $E_0 - E_F$, $m^*$ and the g-factor.[17] The small shift of 0.015 $T^{-1}$ of our results from the calculated (inverse) field values for the extrema is caused by quantum confinement; $m^*$ is observed not to be modified.

The d(MR)/dB peak-sequences indicated by S and BL in Figure 2, are the main motivation for the present letter. These MR peaks are not observed for bulk Bi but are strong in nanowires which suggest a surface origin. As is shown by the polar plots of the SdH periods [Figure 2], 30-nm nanowires show an isotropic period. In contrast, in the 50-nm wires the periods for both the S-and BL-sequences, is very anisotropic. This is shown in Fig 2; i.e., the LMR does not show it but the MR for $\theta = 15^\circ$ does. Regarding the S-sequence in 200-nm nanowires, we only show the LMR results but the same sequence is observed in the TMR and with the same period. The behavior exhibited by the S and BL sequences, periodicity in $1/H$, and monotonic angular dependence consistent with a convex Fermi surface, is associated with 3D Landau levels. Now we analyze the data to extract effective mass and charge density of S and BL sequences. The values of the effective mass are obtained from the temperature dependence of the SdH amplitudes. For 30-nm nanowires, we get $m^* \sim 0.3\ m_e$, and $m^* \sim 0.06\ m_e$, for the carriers that are involved in the S-sequence and BL-sequence, respectively. The observations of effective mass support a "surface states" interpretation for the S-sequence, because of its high value close to $m_\Sigma$, and a quantum confinement of bulk states interpretation for the BL sequence because of its low value, close to $m^*$ for bulk Bi. Regarding charge density, for a bulk, macroscopic, single crystal Bi, $\Delta(1/B)_{h,//} = 0.157\ \text{T}^{-1}$ ($\theta=0$), this is the sequence that contains h1⁻ and h2⁻ peaks, and $\Delta(1/H)_{h,\perp} = 0.045\ \text{T}^{-1}$ ($\theta = 90^\circ$). Therefore, one finds finds $p = 3.0 \times 10^{17}/\text{cm}^3$. For our 200-nm nanowires, where $\Delta(1/H)_{h,//} = 0.122\ \text{T}^{-1}$ and $\Delta(1/H)_{h,\perp} = 0.048\ \text{T}^{-1}$ we find that $p = 3.15 \pm 0.3 \times 10^{17}/\text{cm}^3$. Therefore, within the margin of error, $p$ is not modified by quantum confinement in the 200-nm nanowires, even though the actual peak positions have changed. 30-nm Bi array samples show isotropic S- and BL- SdH periods of 0.025 T⁻¹ and 0.13 T⁻¹, respectively. This is consistent with spherical pockets of the Fermi surface of density $n(S) = (1.3 \pm 0.2) \times$ $10^{18}$/cm$^3$ and n(BL) = (1.0 ± 0.5) × $10^{17}$/cm$^3$. Therefore we find $n$(BL) ~1/3 $n_0$ even though, for 30-nm nanowires, $E_c > E_0$ ; we believe this is an effect akin to that one observed by Hirahara et al [13] and that was interpreted in terms of hybridization between surface and bulk-like states. Regarding the S-sequences, SdH orbits extend over the Larmor radius $r_L$ which is $m^* m_0 V_F / |e| H$. For these Landau states, for $H$ = 5 T, we get $r_L$ =17 nm. Therefore the surface states in 30 nm nanowires fill a significant volume fraction of the 30 nm diameter wires, and therefore cannot be considered a 2D cylindrical conductor covered by surface charges. However, the value of the surface charge per unit area $\sigma$ can be estimated as $n(S) \times r_L$ and we find $\sigma$ = 2.2 × $10^{12}$/cm$^2$ which is in rough agreement with the $\Sigma$ as measured by ARPES. The existence of BL states and the observation of a large anisotropy in the 50-nm case where $\lambda_F \sim d$ is consistent with the observations, using infrared absorption spectroscopy, of low-effective mass electrons and holes molded by quantum confinement including quantum interference around the wire.[18]

    Even if the nanowire is a single crystal, the surface cannot be a single low-index crystal of the type that is studied in ARPES experiments but rather it consists of a finite set of different surfaces as one goes around. A number of these surfaces were studied using ARPES and this method shows that surface states with relatively high effective masses and similar carrier densities are present on all low-index surfaces studied so far. ARPES measurements show that the surface states are essentially 2D, with an anisotropic extent in the direction perpendicular to the surface often called penetration length. The surface states of some of the semi-infinite surfaces of the orientations perpendicular to

the trigonal penetrate deeply into the bulk of the crystal in contrast to, for example the trigonal, [0001], surfaces where the surface states penetrate little. Our nanowire correspond to the case of large penetration length.

Consistent with our identification of the observed resonances with 3D Landau level crossings through the Fermi level, we index the S and BL peaks by integers $N_S$ and $N_{LB}$ corresponding to the Landau states ocupation number. Figure 3 shows such a plot of index versus inverse field location and shows that the data is consistent with the expectation of constant slopes.

The existence of surface carriers is well-established, as evidenced by the work presented here and ARPES measurements. Can these carriers be related to the unusual thermopower and Hall data presented in Refs. 3 and 4? The thermopower extrema from Ref. 3 are also presented in Figure 3. Remarkably we find that the nanowires data for the S sequences and the thermopower results are similar in slope, strongly suggesting a common origin. Our data extends only to 0.07 $T^{-1}$. Assuming that the data from Ref. 3 arise from surface states, this range of observation is smoothly extended to 0.03 $T^{-1}$. The slope changes at high magnetic fields because the Fermi level is not stationary near a quantum limit. The true quantum limit is reached when the extrapolation of data in Figure 3 crosses the absisa and this can be estimated to be 0.015 $T^{-1}$ or 70 T. It is only for these high magnetic fields that we expect filling factors of the order of 1 and the rich physics that is anticipated by theory is realized. The quantum limit for holes is found to be 8.3 T and in this case the indexing is absolute. In contrast, the indexing that we have made of the S peaks is tentative, it can be changed by an integer typically indexing by 0 the last

LS level to dissapear as we ramp up the field. Clearly the available data does not allow us to make an absolute indexing in this case.

The surface states described here are not evident in bulk electronic transport measurements. However, the measurements presented in Refs. 3 and 4, a variation on thermopower measurements and Hall effect, respectively, are quite sensitive to surfaces, and so we believe that our alternative interpretation of their results in terms of hybridized surface states is plausible.

To summarize, the electronic structure and transport in small diameter trigonal Bi nanowires (30 nm < diameter< 200 nm), was investigated. Surfaces of Bi support mobile charge carriers because of the spin-orbit interaction associated with the surface electric field. The surface electronic conduction band is metallic, in contrast with the electron and hole bands in the semimetallic bulk. We find that in trigonal nanowires, surface carriers exhibit three dimensional behavior through a rich spectrum of Landau levels in an spherical Fermi surface possibly because their wavefunction penetrates deeply in the interior of Bi. Small diameter nanowires mirror the behavior of the boundary of macroscopic Bi samples in the purported 'quantum limit' *of H* > 9 T [3,4] because the holes' energy relaxation mechanism through surface states, that we have observed in nanowires, explains these recent thermopower (Nernst) measurements. These observations have been interpreted as a signature of charge fractionalization in bulk Bi at high magnetic fields, and we propose an interpretation in terms of surface states. The quantum limit for surface states is found to be 70 T.

**FIGURE CAPTIONS**

Figure 1.    Fermi surfaces of the bulk and the boundary layer of bismuth. Bulk: It shows the spatial distribution, with respect to the Brillouin zone, of the wavevectors of L-electrons and T-holes whose energies are at the Fermi level. The crystalline orientation of the trigonal nanowires is shown. In these nanowires, the three electron pockets are crystallographically equivalent. Boundary layer: the spherical Fermi surface of the boundary charges demonstrated in the present paper.

Figure 2.    Top: Angle-dependent derivative of the magnetoresistance of Bi wire array samples of various diameters as indicated. Inset: Schematic view of the nanowire array sample with silver epoxy contact; the magnetic field is represented as an arrow at an angle $\theta$ with respect to the wirelengths. Bottom: Polar plot of the SdH periods of holes (circles) in bulk Bi and assigned to S (squares) and BL (circles) quasiparticles in 200-nm, 50-nm and 30-nm nanowire samples with a guide for the eye, the dotted arc, where appropriate. The symmetry, spherical or ellipsoidal, of these plots reflects the symmetry of the corresponding Fermi surfaces.

Figure 3.    Circles. Points in Behnia *et al* (Science, <u>317</u> 1729 (2007)) that were interpreted in terms of fractionalization; the fractions are indicated. Instead of designating the points as fractional we assign them a sequential

integral index $N_S$. The dashed line shows the extrapolation that gives a surface states quantum limit of 70 T. Up-triangles: Boundary Landau states data of 30-nm Bi nanowires. Inset: Bulk-like Landau states data (down-triangles) of 30-nm nanowires.

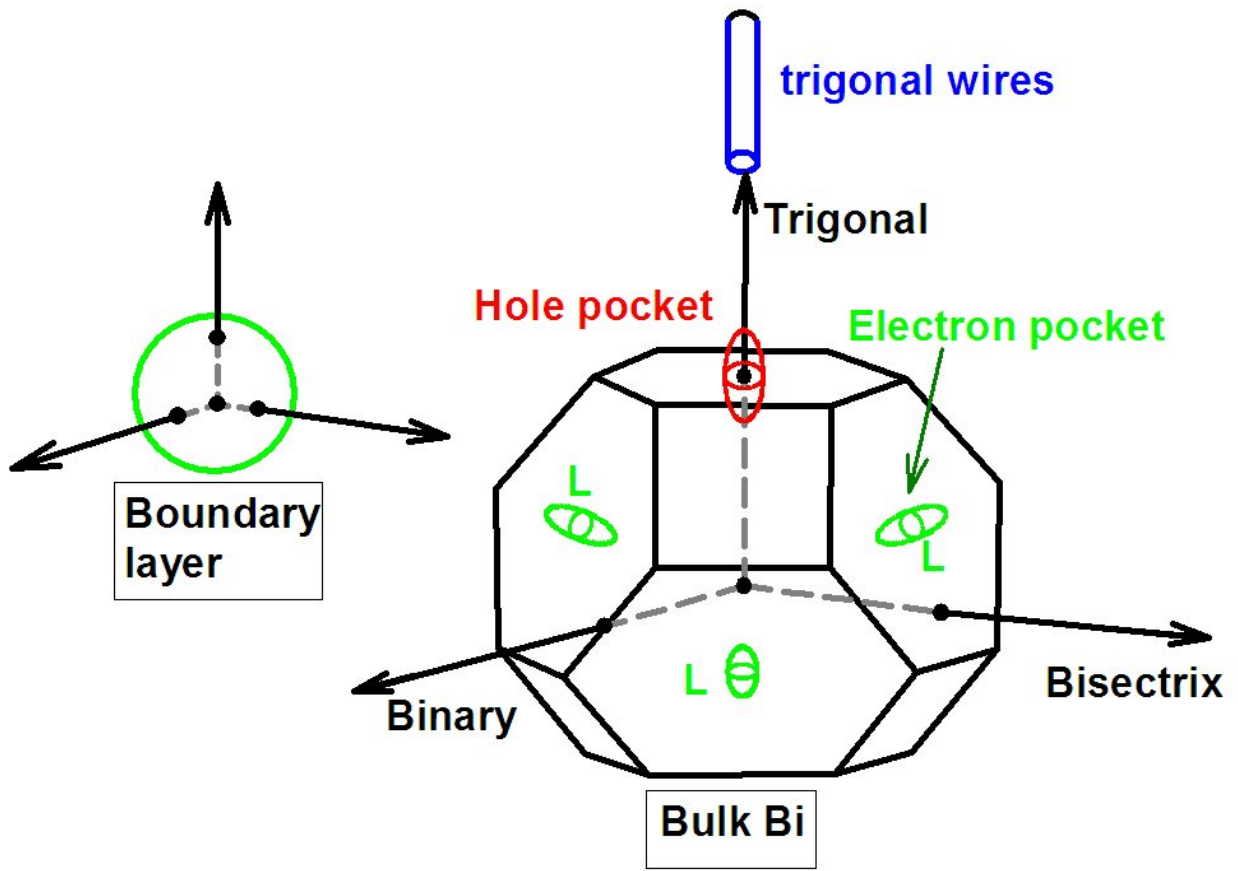

Figure 1.

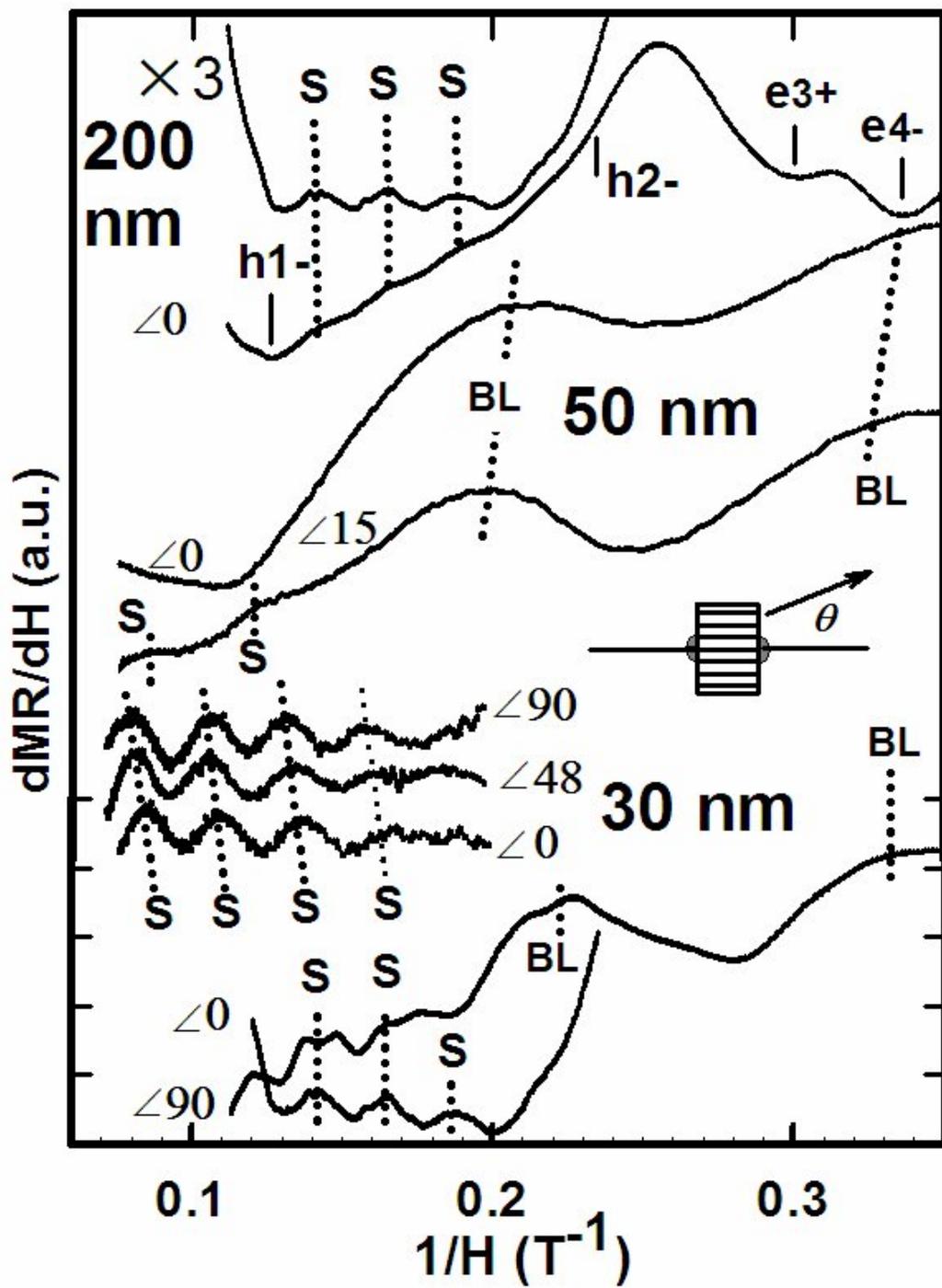

Figure 2. Top.

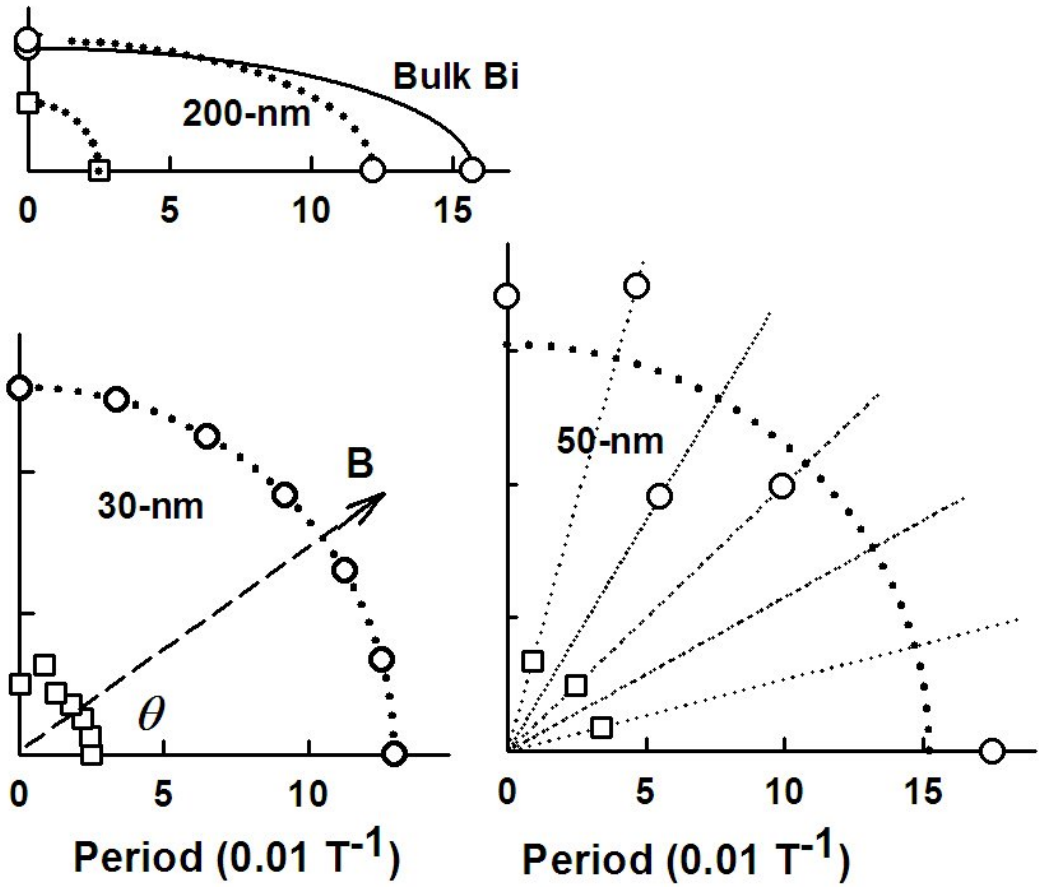

Figure 2. Bottom.

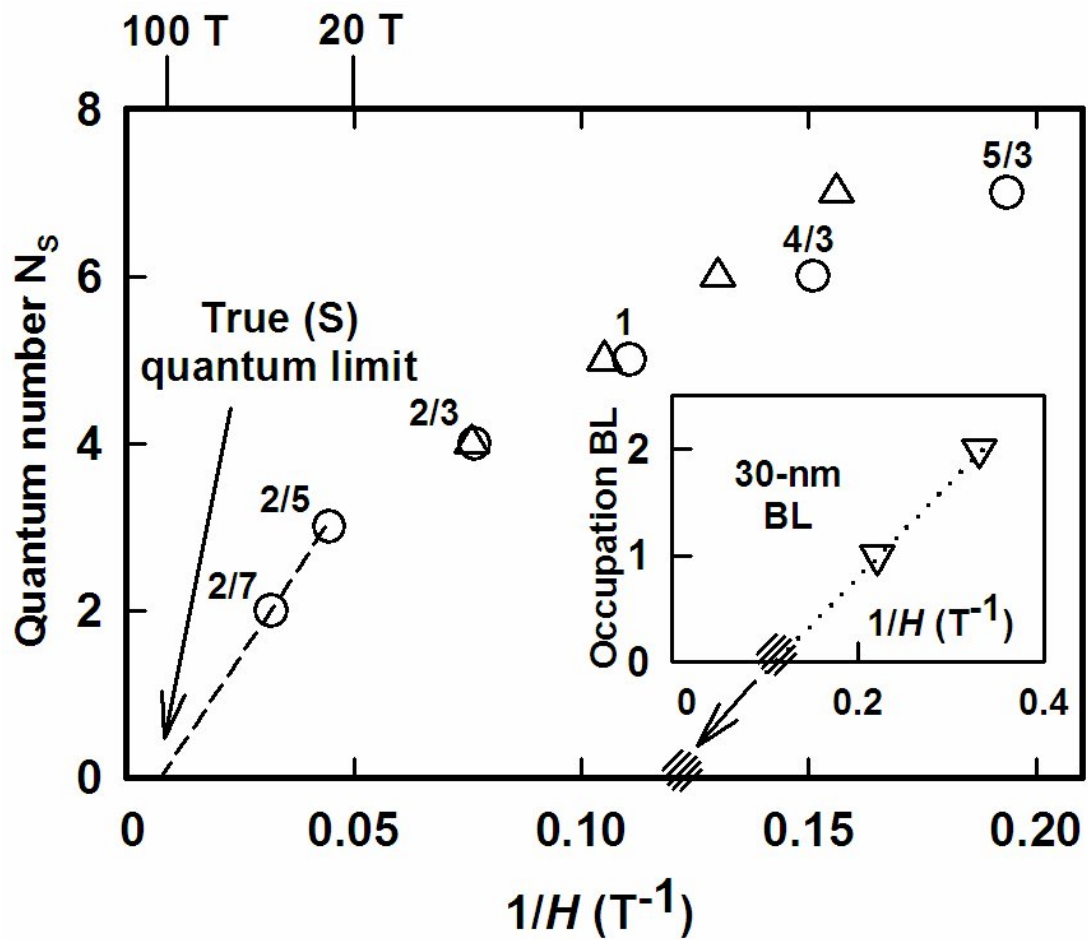

Figure 3.